\documentclass[conference]{IEEEtran}

\usepackage[colorlinks=true, linkcolor=blue, citecolor=blue, urlcolor=blue]{hyperref}

\usepackage{url}
\usepackage{amsmath}
\usepackage{graphicx}
\usepackage{subcaption}
\usepackage{booktabs}
\usepackage{tabularx}
\usepackage{booktabs}
\usepackage{balance}
\usepackage{tcolorbox}

\newcommand{\E}{\textmd{E}}
\newcommand{\Ga}{\textmd{G}}
\newcommand{\Var}{\textmd{Var}}
\newcommand{\Pois}{\textmd{Pois}}

\usepackage{fancyhdr} 
\fancypagestyle{firststyle}
{
   \fancyhf{}
   \fancyfoot[C]{\scriptsize{Proceedings of the 18th Annual International Conference on Privacy, Security and Trust (PST 2021), Auckland (online), IEEE, pp.~1--10. This is the authors' copy. The publisher's copy is available online via \url{https://doi.org/10.1109/PST52912.2021.9647791}}}
   }

\begin{document}

\title{A Large-Scale Security-Oriented Static \\ Analysis of Python Packages in
  PyPI}


\author{
\IEEEauthorblockN{Jukka Ruohonen}
\IEEEauthorblockA{University of Turku, Finland \\
Email: juanruo@utu.fi}
\and
\IEEEauthorblockN{Kalle Hjerppe}
\IEEEauthorblockA{University of Turku, Finland \\
Email: kphjer@utu.fi}
\and
\IEEEauthorblockN{Kalle Rindell}
\IEEEauthorblockA{University of Turku, Finland \\
Email: kakrind@utu.fi}
}

\maketitle

\begin{abstract}
Different security issues are a common problem for open source packages archived to and delivered through software ecosystems. These often manifest themselves as software weaknesses that may lead to concrete software vulnerabilities. This paper examines various security issues in Python packages with static analysis. The dataset is based on a snapshot of all packages stored to the Python Package Index (PyPI). In total, over 197 thousand packages and over 749 thousand security issues are covered. Even under the constraints imposed by static analysis, (a)~the results indicate prevalence of security issues; at least one issue is present for about 46\% of the Python packages. In terms of the issue types, (b)~exception handling and different code injections have been the most common issues. The \texttt{subprocess} module stands out in this regard. Reflecting the generally small size of the packages, (c)~software size metrics do not predict well the amount of issues revealed through static analysis. With these results and the accompanying discussion, the paper contributes to the field of large-scale empirical studies for better understanding security problems in software ecosystems.
\end{abstract}


\maketitle

\section{Introduction}

\thispagestyle{firststyle} 

Python is currently among the most popular programming languages. Like many of
today's poplar programming languages, Python also provides its own ecosystem for
archiving and maintaining open source software packages written with the
language. This ecosystem is known as the Python Package Index. Recently, this
index and related language-specific repositories have been extensively studied
from different angles. Maintenance, release engineering, and dependencies have
provided typical motivations for the studies. Security has provided a further
motivation. All motivations intersect with methodological questions in empirical
software engineering.

This intersection also provides the reference point for the present work; the
paper continues the recent large-scale empirical analyses on software security
issues in software ecosystems~\text{\cite{Korkmaz20, LiuMeng20,
    Zimermann19}}. As is elaborated in the opening Section~\ref{sec: research
  question}, the paper also belongs to a specific corner in this research
branch: the packages within PyPI are analyzed in isolation of each other; the
ecosystem concept is understood in statistical terms as a population rather than
in technical terms as a collection of more or less interlinked packages. As is
further elaborated in the section, the paper can be further framed with respect
to the method of detecting (but not verifying) different security ``issues'',
which may, or may not, equate to actual security~bugs. Often, the term ``smell''
is used as an alternative.

In other words, the paper is closely tied to static analysis, which has long
been used as an alternative to other means to discover security issues during
software developing, including security-related code reviews (see
\cite{Shahriar12} for a comprehensive review of the history and associated
literature). However, neither static analysis nor code reviews are sufficient
alone; both tend to miss many security issues~\cite{Edmundson13}. In addition to
discussing this point in more detail, the opening section notes the framing
toward the Python programming language itself. Afterwards, the structure is
fairly straightforward: the large-scale empirical approach is outlined
Section~\ref{sec: materials and methods} together with the statistical methods
used; results are presented in Section~\ref{sec: results} and further discussed
in \ref{sec: discussion} in conjunction with their limitations; and a brief
conclusion follows in the final Section~\ref{sec: conclusion}.

\section{Research Design}\label{sec: research question}

\subsection{Research Questions}\label{subsec: research questions}

The following three research questions (RQs) are examined:
\begin{itemize}
\itemsep 3pt
\item{$\textmd{RQ}_1$: \textit{How common and severe are security issues in PyPI packages given constraints of static program analysis?}}
\item{$\textmd{RQ}_2$: \textit{What types of security issues are typical to PyPI packages given again the constraints of static analysis?}}
\item{$\textmd{RQ}_3$: \textit{What is the average number of security-related issues when the size of the packages is controlled for?}}
\end{itemize}

These questions are easy to briefly justify. Given the ecosystem-wide focus soon described, $\textmd{RQ}_1$ is worth asking because the empirical dataset covers the whole PyPI. Bug severity is also a classical topic in software security research~\cite{Ruohonen19ACI, Spanos18}. It~is further related to the types of security issues detected~($\textmd{RQ}_2$), which are always interesting and relevant for making practical improvements. Although all three questions are constrained by the static analysis tool used, these constraints are balanced by the large-scale analysis. That is, generalizability is sought toward all packages stored to PyPI but not toward all static analysis tools---let alone all security issues. This framing of the paper's scope is a typical ``pick one'' choice often encountered in practical software engineering: not all generalizability types can be achieved in a single study under reasonable time and resource constraints.

The final $\textmd{RQ}_3$ can be motivated with existing literature on the relation between software vulnerabilities and software metrics. A basic finding in this literature is that different software metrics (such as those related to quality and design) correlate with the amount of vulnerabilities---yet these metrics correlate also with the size of software~\cite{Alves16, Chowdhury11}. These correlations make interpretation less straightforward. Although numerical indicators on software quality correlate with the amount of vulnerabilities discovered (and, to some extent, can even predict these), it may be that the underlying theoretical dimension is merely software size, or that software size is a confounding factor~\cite{Fenton99}. Therefore, it has been argued that when predicting bugs, vulnerabilities, weaknesses, smells, or other software quality attributes, software size metrics are mainly useful as covariates (a.k.a.~control variables)~\cite{Kitchenham10}. To this end, concepts such as vulnerability density have also been introduced to gain more robust empirical measurements.

Although scaling by software size (such as the lines of code) has been common~(see~\cite{Ruohonen19EASE} and references therein), scaling can be done also in terms of testing coverage and continuous integration traces~\cite{Gkortzis18}. Static analysis coverage is used in the present work; $\textmd{RQ}_3$ is answered by using the number of lines and files of the Python code scanned by the static analysis tool used. The gist behind  $\textmd{RQ}_3$ is simple: if the results indicate that the average number of issues is similar with and without conditioning by these software size metrics, there is a little rationale to use vulnerability densities or related scaling. This reasoning depends on the context; Python and the PyPI software ecosystem, static analysis, and security issues.

\subsection{Related Work}

Instead of enumerating individual papers, the paper's focus is better elaborated by considering four large branches of related research. These branches allow to also further frame the paper's scope. The branches can be summarized as follows.

\subsubsection{Software Ecosystems}

Software ecosystems cover a large research branch in software engineering, information systems research, and related fields. According to a famous definition, a ``software ecosystem is a collection of software projects which are developed and evolve together in the same environment''~\cite[p.~265]{Lungu10}. On the one hand, this definition underlines the presence of a common environment, which is typically orchestrated by a company or a community. Thus, the definition includes commercial software ecosystems (such as Google Play) and ecosystems that focus on providing supplementary functionality to a software framework, such as the WordPress plugin ecosystem~\cite{Ruohonen19EASE}. That said, software package repositories (such as \textit{npm} for JavaScript, CPAN for Perl, or CRAN for R) are likely the most studied software ecosystems in empirical research. Also PyPI has been studied recently \text{\cite{Korkmaz20, Ruohonen18IWESEP, Valiev18}}. In light of the definition, the Python Package Index indeed provides a common environment, although the actual development mainly occurs on GitHub, another software ecosystem. On the other hand, the definition underlines common software evolution, which, in turn, pinpoints toward dependencies between packages and longitudinal analysis. For framing the present work, it is important to underline that neither apply: the empirical analysis is based on a snapshot and the packages are analyzed in isolation from each other. Finally, it should be remarked that only publicly available open source packages are considered; PyPI is the theoretical population.

\subsubsection{Static Analysis}

Static analysis covers a huge research branch in software engineering and computer science in general. Typical application domains relate to portability, coding style enforcement, reliability, and maintainability~\cite{Ogasawara98}. According to this high-level framing, security-oriented static analysis is a subset of the reliability domain. Within this subset, static analysis is typically used to either explicitly detect software vulnerabilities, or to provide warnings to developers about unsafe programming practices that may lead to different security issues, including software weaknesses that may manifest themselves as concrete software vulnerabilities. The warning-related use case applies to the present work; none of the issues in the empirical analysis have been explicitly verified to have security implications, although some implicit confidence exists that some of the issues truly are security issues. In terms of static analysis research, further framing can be done with a distinction to technical and social research topics. In general, the former deal with tools for specific programming languages, while the latter studies the question of how these tools are used with particular languages~\text{\cite{Beller16, Nachtigall19, Vassallo20}}. The social topics include also the further question of how developers use static analysis tools to diagnose and fix weaknesses and vulnerabilities~\text{\cite{Oyetoyan18, Smith19}}. Although the paper does not present a new tool, the topic is still technical rather than social in the sense that nothing is said about the development and developers of the Python~packages.

\subsubsection{Security Issues}

The third related research branch covers the detection of different security issues during software testing. The branch is again large---to say the least. Even when limiting the scope to static analysis, numerous relevant questions are present: which issues are detected; how tools differ for a given programming language; how detection differs between languages; how fault-injection compares to signature-based approaches and how these differ from mutation analysis and search-based analysis; and so on~\text{\cite{Shahriar12, Aloraini17}}. These technical questions can be again augmented by social ones; how detection based on static analysis fits into software testing in general; how effective is such detection in terms of time and resource constraints; and so forth. In terms of security issues themselves, typically either known vulnerabilities, usually based on the Common Vulnerabilities and Exposures (CVEs), or the abstract weaknesses behind these, often based on the corresponding Common Weakness Enumeration (CWE) framework, are used on both sides of the socio-technical research paradigm~\text{\cite{Oyetoyan18, Benabidallah19}}. Roughly, CVEs are more useful on the technical side and CWEs on the social side of research. For instance, CWEs have been used to better understand typical programming mistakes in Python packages~\cite{Ruohonen18IWESEP}, to provide dynamic information sources for software developers using static analysis tools~\cite{Smith19, RedHat20a}, and so forth. Despite the advantages of such approaches also for systematic empirical analysis, the paper's approach is based on the issue categories provided by the static analysis tool used. Although the issues caught by the tool can be mapped to CWEs~\cite{Rahman19}, such mapping yields less fine-grained categories and must be done manually.


\subsubsection{Python}

The Python programming language itself constitutes the fourth and final branch of related research. This branch intersects with all previous three branches. As already remarked, PyPI has been studied in previous work, but there are also many Python-specific studies on security issues and static analysis. Regarding the latter, it is necessary to point out the obvious: Python is a dynamically typed language, which, on one hand, makes static analysis tools less straightforward to implement. On the other, the typing system also makes static analysis tools particularly valuable for Python and related languages~\cite{Beller16}. Against this backdrop, it is no surprise that many static analysis tools have been developed for Python in both academia and industry. These range from simple linter-like checkers (such as the one used in the present work) to formal verification methods and actual static type systems~\cite{Fromherz18}. In addition, many related datasets and empirical studies have been conducted, including studies focusing on traditional software metrics (such as those related to code complexity and object-orientation)~\cite{Orru15}, code smells \cite{Zhifei18}, call graphs~\cite{Li19a}, and the Python's type system \cite{Xia18}. Within this empirical research domain, there exists also one previous study that shares the same motivation and uses the same static analysis tool:~\cite{Rahman19}. As will be elaborated in the subsequent section, the datasets, their operationalization, and methods are still very~different.

\section{Materials and Methods}\label{sec: materials and methods}

\subsection{Data}

\begin{figure}[th!b]
\centering
\includegraphics[width=7.5cm, height=1.2cm]{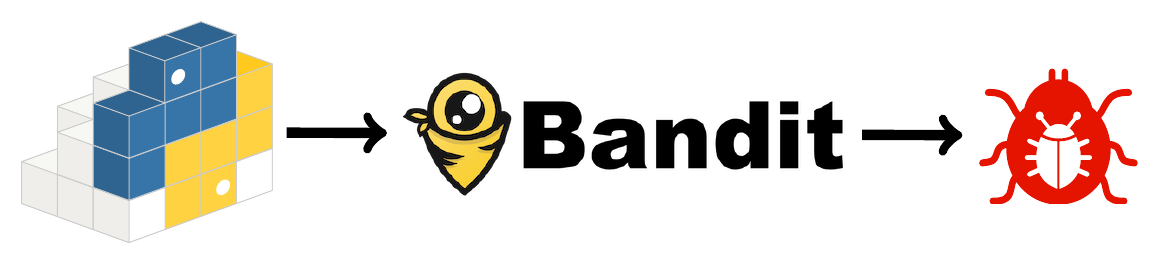}
\caption{Sample Construction in a Nutshell}
\label{fig: sample}
\end{figure}

The dataset is based on a simple index file provided in the Python Package Index~\cite{PyPI20a}. In total, $224,651$ packages were listed in the index at the time of retrieving it. Given these packages, the most recent releases were downloaded from PyPI with the \texttt{pip} package manager using the command line arguments \texttt{download --no-deps}. Because of the dynamic index file, (a)~it should be remarked that some packages were no longer available for download. Furthermore, (b)~only those packages were included that were delivered as well-known archive files (namely, as \textit{tar}, \textit{gzip}, \textit{bzip2}, \textit{xz}, ZIP, or RAR files). After extracting the archives, the contents were fed to the Bandit~\cite{bandit} static analysis tool using Python 3.6.10. This feeding is illustrated in Fig.~\ref{fig: sample}. Given the ongoing painful transition from Python~2 to Python~3~\cite{Malloy19}, (c)~those packages had to be excluded that were compatible only with Python~2.7. Finally, (d) those packages were also excluded for which Bandit reported having not scanned a single line or a file.

Despite the necessary exclusions discussed, the dataset constructed contains the
static analysis results for as many~as $n = 197,726$ packages. Thus, the size of
the dataset is very similar to other recent large-scale studies (for instance,
$192,666$ Python packages were retrieved from PyPI in~\cite{Korkmaz20}). A brief
remark is also in order about the tool providing the results in the present
work. This static analysis tool was originally developed by Hewlett-Packard and
associates for use in the OpenStack project. It has been used also in previous
research~\cite{Rahman19}. Given existing taxonomies for static analysis
tools~\text{\cite{ChessMcGraw04, Hjerppe19IFIP}}, Bandit exploits the
conventional abstract syntax tree but operates mainly at the local level;
statements, functions, and parameters to functions are parsed without
considering the flow and semantics between these. The tool's approach can be
further elaborated by considering the issues.

\subsection{Issues}

Bandit contains several individual detectors that can be enabled or disabled
according to a project's specific needs. Given the PyPI-wide focus, all of these
were included in the sample construction except a detector for the use of
assertions~(B101), which was prone to false positives according to preliminary
tests. The individual issue types detected are enumerated in Table~\ref{tab:
  issues}. The three first columns in the table display identification codes for
the detectors, their mnemonic names, and brief descriptions for the issues
detected. The fourth and fifth columns contain Bandit's metrics for the
\textit{severity} (S) of the issues detected and the \textit{confidence} (C) of
detection for a given issue. Both are important in practice. Static analysis
tools are highly prone to false negatives (bugs not caught) and false positives
(issues that are not bugs)~\text{\cite{ChessMcGraw04, Edmundson13}}. Including
detection confidence indicators is one way for addressing the
problem~\cite{Nachtigall19}. Analogously, according to recent surveys, issue
severity is the most important factor for software developers in their
prioritization tasks~\cite{Vassallo20}. The severity and confidence metrics take
three values: \textit{low}~(L), \textit{medium}~(M), and
\textit{high}~(H). Although there is a long-standing debate about ranking the
severity of security-related issues, including actual vulnerabilities in
particular~\cite{Spanos18, Spring18a}, the values provided in Bandit are taken
for granted in order to ensure replicability. Finally, the table's last column
provides references to further information about the issues. When available,
these references again conform with those given in the Bandit's source~code.

\begin{table*}[t!]
\centering
\begin{scriptsize}
\caption{Issue Types Detected by Bandit (excluding assertions)}
\label{tab: issues}
\begin{tabular}{lllllr}
\toprule
Code & Mnemonic name & Description & S & C & Ref. \\
\hline
B102 & \texttt{exec\_used} & Use of the \texttt{exec} function & M & H & \cite{B102} \\
B103 & \texttt{set\_bad\_file\_permissions} & Insecure permissions for files & M, H & M, H & \cite{B103} \\
B104 & \texttt{hardcoded\_bind\_all\_interfaces} & Binding a socket to all network interfaces & M & M & \cite{B104} \\
B105 & \texttt{hardcoded\_password\_string} & Use of hard-coded passwords in non-function contexts & L & L & \cite{B105} \\
B106 & \texttt{hardcoded\_password\_funcarg} & Use of hard-coded passwords in function arguments & L & M & \cite{B105} \\
B107 & \texttt{hardcoded\_password\_default} & Use of hard-coded passwords in default function arguments & L & M & \cite{B105} \\
B108 & \texttt{hardcoded\_tmp\_directory} & Use of hard-coded temporary directories & M & M & \cite{B108} \\
B110 & \texttt{try\_except\_pass} & Using \texttt{pass} as a catch-all-style exception handling & L & H & \cite{B110} \\
B112 & \texttt{try\_except\_continue} & Using \texttt{continue} as a catch-all-style exception handling & L & H & \cite{B110} \\
\hline
B201 & \texttt{flask\_debug\_true} & Running a Flask web application in debug mode & H & H & \cite{B201} \\
\hline
B301 & \texttt{pickle} & Use of insecure deserialization & M & H & \cite{B301} \\
B302 & \texttt{marshal} & Use of insecure deserialization & M & H & \cite{B301} \\
B303 & \texttt{md5} & Use of MD2, MD4, MD5, or SHA1 hash functions & M & H & \cite{B303} \\
B304 & \texttt{ciphers} & Use of insecure ciphers such as DES & H & H & \cite{B303} \\
B305 & \texttt{cipher\_modes} & Use of insecure cipher modes & M & H & \cite{B303} \\
B306 & \texttt{mktemp\_q} & Use of the insecure \texttt{mktemp} function & M & H & \cite{B306} \\
B307 & \texttt{eval} & Use of the possibly insecure \texttt{eval} function & M & H & \cite{B307} \\
B308 & \texttt{mark\_safe} & Use of the possibly insecure \texttt{mark\_safe} function & M & H & \cite{B308} \\
B309 & \texttt{httpsconnection} & Use of the insecure \texttt{HTTPSConnection} with some Python versions & M & H & \cite{B309} \\
B310 & \texttt{urllib\_urlopen} & Use of a file scheme in \texttt{urlopen} with some Python versions & M & H & \cite{B310} \\
B311 & \texttt{random} & Use of pseudo-random generators for cryptography/security tasks & L & H & \cite{B311} \\
B312 & \texttt{telnetlib} & Use of the insecure Telnet protocol & H & H & \cite{B312} \\
B313 & \texttt{xml\_bad\_cElementTree} & Use of possibly insecure Extensible Markup Language (XML) parsing & M & H & \cite{B313} \\
B314 & \texttt{xml\_bad\_ElementTree} & Use of possibly insecure XML parsing & M & H & \cite{B313} \\
B315 & \texttt{xml\_bad\_expatreader} & Use of possibly insecure XML parsing & M & H & \cite{B313} \\
B316 & \texttt{xml\_bad\_expatbuilder} & Use of possibly insecure XML parsing & M & H & \cite{B313} \\
B317 & \texttt{xml\_bad\_sax} & Use of possibly insecure XML parsing & M & H & \cite{B313} \\
B318 & \texttt{xml\_bad\_minidom} & Use of possibly insecure XML parsing & M & H & \cite{B313} \\
B319 & \texttt{xml\_bad\_pulldom} & Use of possibly insecure XML parsing & M & H & \cite{B313} \\
B320 & \texttt{xml\_bad\_etree} & Use of possibly insecure XML parsing & M & H & \cite{B313} \\
B321 & \texttt{ftplib} & Use of the clear-text sign-in File Transfer Protocol (FTP) & H & H & \cite{B312} \\
B322 & \texttt{input} & Use of the insecure \texttt{input} function (with Python 2) & H & H & \cite{B307} \\
B323 & \texttt{unverified\_context} & Explicitly bypassing default certificate validation & M & H & \cite{B323} \\
B324 & \texttt{hashlib\_new\_insecure\_functions} & Use of MD2, MD4, MD5, or SHA1 hash functions with \texttt{hashlib} & M & H & \cite{B303} \\
B325 & \texttt{tempnam} & Use of the insecure and deprecated \texttt{tempnam} or \texttt{tmpnam} functions & M & H & \cite{B325} \\
\hline
B401 & \texttt{import\_telnetlib} & Import of a Telnet library & H & H & \cite{B312} \\
B402 & \texttt{import\_ftplib} & Import of a FTP library & H & H & \cite{B312} \\
B403 & \texttt{import\_pickle} & Import of a library for deserialization & L & H & \cite{B301} \\
B404 & \texttt{import\_subprocess} & Import of the possibly insecure \texttt{subprocess} library & L & H & \cite{B404} \\
B405 & \texttt{import\_xml\_etree} & Import of a possibly insecure XML parsing library & L & H & \cite{B313} \\
B406 & \texttt{import\_xml\_sax} & Import of a possibly insecure XML parsing library & L & H & \cite{B313} \\
B407 & \texttt{import\_xml\_expat} & Import of a possibly insecure XML parsing library & L & H & \cite{B313}  \\
B408 & \texttt{import\_xml\_minidom} & Import of a possibly insecure XML parsing library & L & H & \cite{B313} \\
B409 & \texttt{import\_xml\_pulldom} & Import of a possibly insecure XML parsing library & L & H & \cite{B313} \\
B410 & \texttt{import\_lxml} & Import of a possibly insecure XML parsing library & L & H & \cite{B313} \\
B411 & \texttt{import\_xmlrpclib} & Import of a possibly insecure XML parsing library & H & H & \cite{B313} \\
B412 & \texttt{import\_httpoxy} & Exposition to the so-called ``httpoxy'' vulnerabilities & H & H & \cite{B412} \\
B413 & \texttt{import\_pycrypto} & Use of the deprecated Python Cryptography Toolkit (pycrypto) & H & H & \cite{B413}\\
\hline
B501 & \texttt{request\_with\_no\_cert\_validation} & Ignoring the validation of certificates & H & H & \cite{B501} \\
B502 & \texttt{ssl\_with\_bad\_version} & Use of old and insecure Transport Layer Security (TLS) versions & H & H & \cite{B502} \\
B503 & \texttt{ssl\_with\_bad\_defaults} & Use of default parameter values that may yield insecure TLS transports & M & M & \cite{B502} \\
B504 & \texttt{ssl\_with\_no\_version} & Use of default parameter values that permit insecure TLS transports & L & M & \cite{B502}  \\
B505 & \texttt{weak\_cryptographic\_key} & Using  inadequate key length for a cipher & H & H & \cite{B303} \\
B506 & \texttt{yaml\_load} & Insecure use of the \texttt{load} function from the PyYAML library & M & H & \cite{B506} \\
B507 & \texttt{ssh\_no\_host\_key\_verification} & Not verifying keys with a Python library for
Secure Shell (SSH) & H & M & \cite{B507} \\
\hline
B601 & \texttt{paramiko\_calls} & Exposure to command injection via a Python library for SSH & M & M & \cite{B601} \\
B602 & \texttt{subprocess\_popen\_with\_shell\_equals\_true} & Spawning a \texttt{subprocess} using a command shell & L, M, H & H, H, H & \cite{B602} \\
B603 & \texttt{subprocess\_without\_shell\_equals\_true} & Spawning a \texttt{subprocess} without a shell but without input validation & L & H & \cite{B603} \\
B604 & \texttt{any\_other\_function\_with\_shell\_equals\_true}  & Using a wrapper method with a command shell (excluding B602) & M & H & \cite{B601} \\
B605 & \texttt{start\_process\_with\_a\_shell} & Exposure to command injection with a shell (excluding B602 and B604) & L & M & \cite{B602} \\
B606 & \texttt{start\_process\_with\_no\_shell} & Exposure to command injection without a shell (excluding B603) & L & M & \cite{B601} \\
B607 & \texttt{start\_process\_with\_partial\_path} & Spawning a process without absolute path & L & H & \cite{B601} \\
B608 & \texttt{hardcoded\_sql\_expressions} & Exposure to Structured Query Language (SQL) injection & M & L & \cite{B608} \\
B609 & \texttt{linux\_commands\_wildcard\_injection} & Exposure to command injection via Unix wildcards & H & M & \cite{B609} \\
B610 & \texttt{django\_extra\_used} & Exposure to SQL injection via the Django framework & M & M & \cite{B608} \\
B611 & \texttt{django\_rawsql\_used} & Exposure to SQL injection via the Django framework & M & M & \cite{B608} \\
\hline
B701 & \texttt{jinja2\_autoescape\_false} & Exposure to Cross-Site Scripting (XSS) via a Python library & H, H & H, H & \cite{B701} \\
B702 & \texttt{use\_of\_mako\_templates} & Exposure to XSS via a Python library & M & H & \cite{B701} \\
B703 & \texttt{django\_mark\_safe} & Exposure to XSS via a Python library & M & H & \cite{B701} \\
\bottomrule
\end{tabular}
\end{scriptsize}
\end{table*}

The issues can be grouped into seven categories. These are:
\begin{enumerate}
\item{The first category contains detectors for \textit{generic} issues that are well-known to be risky. Particularly noteworthy are the heuristic detectors for the presence passwords hard-coded to the source code. Although Bandit's detection confidence is not high for these issues, hard-coded credentials have been behind many recent high-profile data breaches in cloud computing environments~\cite{Borazjani17}.}
\item{The second category includes a single detector for running a particular web application in \textit{debug mode.}}
\item{The third category addresses various \textit{function calls} that may lead to security issues with varying degree of severity. The examples range from the Python's ``pickling'' functionality via insecure cryptography to well-known but unsafe functions in the language's standard library.}
\item{The fourth category is similar to the third, but instead of detection based on individual function calls, additional checks are present based on simpler \textit{import statements}.}
\item{The fifth category (from B501 to B507) contains detectors for issues that relate to insecure \textit{network protocols}. The examples include many well-known security-related issues related to TLS, validation, and authentication.}
\item{The sixth category contains various checks for different \textit{code injections}. These range from conventional SQL injections to code injections via a command line shell.}
\item{The seventh category contains three simple checks for cross-site scripting, which has been a widespread security issue for web applications written in Python~\cite{Ruohonen18IWESEP}.}
\end{enumerate}

Although the seven categories cover plenty of specific and important issues, the list is hardly exhaustive. For instance, issues related to synchronization, unused code, risky numerical values, and resource management are not present. The limited coverage is unfortunate because resource management issues and buffer-related bugs have been even surprisingly common in Python applications~\cite{Ruohonen18IWESEP}. The lack of coverage for numerical issues is also a known problem for many tools \cite{Oyetoyan18}. Though, it is practically impossible to gain a full coverage of the myriad of different security issues in any given tool, whether based on static analysis or something else. There is also a trade-off: covering the various nuts and bolts of a programming language tends to increase false positives and noise in general. Against this backdrop, it is worth remarking that the popular general-purpose \texttt{pylint}~\cite{pylint} was too noisy for the present purposes. The reason relates to the software ecosystem focus: although the various Python Enhancement Proposals (PEPs) cover different best practices, there are literally tens of thousands of individual coding styles used in the hundreds of thousands of packages in PyPI, and so on. All in all, the issues covered in Bandit reflect the particular security requirements and practices in the OpenStack project. Therefore, there are also some special cases (such as B201) that may not be relevant for Python packages in general. Some further limitations are discussed later on in Subsection~\ref{subsec: threats to validity}.

\subsection{Methods}\label{subsec: methods}

Descriptive statistics are used for answering to the first and second research questions. Following related work~\text{\cite{Ruohonen19ACI, Zhifei18}}, the well-documented \cite{DunnSmyth18} negative binomial (NB) regression is used to answer to $\textmd{RQ}_3$. Two NB equations are~examined:
\begin{equation}\label{eq: model 1}
\E(\textit{Issues}_i~\vert~\ln[\textit{Files}_i])
= \exp(\alpha_1 + \beta_1 \ln[\textit{Files}_i]) = \mu_i \end{equation}
and
\begin{equation}\label{eq: model 2}
\E(\textit{Issues}_i~\vert~\ln[\textit{Lines}_i]) =
\exp(\alpha_2 + \beta_2 \ln[\textit{Lines}_i]) = \gamma_i ,
\end{equation}
where $\E(\cdot)$ denotes the expected value, $\textit{Issues}_i$ is the number
of issues detected for the $i$th package, $\alpha_1$ and $\alpha_2$ are
constants, $\beta_1$ and $\beta_2$ are coefficients, $\textit{Files}_i$ and
$\textit{Lines}_i$ are the number of files and lines of code scanned for the
$i$th package, $\mu_i$ and $\gamma_i$ are shorthand notations for the
conditional means, and \text{$i = 1, \ldots, n$}. Given a coefficient
$\varphi_j$, the conditional variances are given by
$\Var(\textit{Issues}_i~\vert~\ln[\textit{Files}_i]) = \mu_i + \varphi_1
\mu^2_i$ and $\Var(\textit{Issues}_i~\vert~\ln[\textit{Lines}_i]) = \gamma_i +
\varphi_2 \gamma^2_i$. Compared to the Poisson regression, overdispersion is
taken into account with the terms $\varphi_1 \mu^2_i$ and $\varphi_2
\gamma^2_i$. Overdispersion is an obvious concern in the present context---by
assumption, there should be plenty of packages for which no security-related
issues were detected. Stated differently, a mixing distribution is~used:
\begin{equation}\label{eq: distributions}
\textit{Issues}_i~\vert~\lambda_i\sim\Pois(\lambda_i)
~\textmd{and}~
\lambda_i\sim\Ga(\delta_i, \varphi_j) ,
\end{equation}
where $\Pois(\cdot)$ denotes the Poisson distribution, $\Ga(\cdot)$ refers to the Gamma distribution, and $\delta_i \in \lbrace \mu_i, \gamma_i \rbrace$. With these two distributional assumptions, it can be further shown that the number of issues follows the negative binomial distribution.

The two software size metrics cannot be included in the same model because these are expectedly highly correlated with each other. Nevertheless, the basic expectation is simple: the estimated $\hat{\beta}_1$ and $\hat{\beta}_2$ should be both positive. Because the amount of code scanned increases more with additional files than with lines of code, it should also hold that $\hat{\beta}_1 > \hat{\beta}_2$. Given that a logarithm is used for $\textit{Files}_i$ and $\textit{Lines}_i$, the difference should not be substantial, however. Otherwise the coefficients are not straightforward to interpret because these are multiplicative with $\hat{\alpha}_1$ and $\hat{\alpha}_2$. For summarizing the results regarding RQ$_3$, averaging is used for the conditional means:
\begin{equation}\label{eq: conditional means}
\overline{\hat{\mu}}
= \frac{1}{n}\sum^n_{i = 1} e^{\hat{\alpha}_1} \textit{Files}_i^{\hat{\beta}_1}
~~~\textmd{and}~~~
\overline{\hat{\gamma}} = \frac{1}{n}\sum^n_{i = 1} e^{\hat{\alpha}_2} \textit{Lines}_i^{\hat{\beta}_2} .
\end{equation}

The equations in \eqref{eq: model 1} and \eqref{eq: model 2} are estimated for (a)~all issues detected, (b) low severity issues, (c) medium severity issues, and (d) high severity issues. Although the estimates for the subset models (b), (c), and (d) are not directly comparable because the sample sizes vary in these severity subsets, the expectations regarding $\hat{\beta}_1$, $\hat{\beta}_2$, and their effects still hold.

Finally, the results are also reported with bootstrapped estimates. Analogously to cross-validation, bootstrapping allows to infer about the accuracy and stability of the regressions (for technical details see \cite{Efron79, Stine89}). To some extent, bootstrapping can be further used to implicitly deduce about whether there is some generalizability toward the larger theoretical population of all open source Python packages. As for computation, the bootstrapping is implemented by drawing $1,000$ random samples (with replacement) containing $10,000$ observations of the dependent and independent variables. For each random sample, the equations \eqref{eq: model 1}  and \eqref{eq: model 2} are then fitted for all four scenarios (a), (b), (c), and (d). The means of the averaged conditional means in \eqref{eq: conditional means} are used for reporting the results.

\section{Results}\label{sec: results}

\subsection{Issues ($\textmd{RQ}_1$)}

\begin{figure}[th!b]
\centering
\includegraphics[width=\linewidth, height=9cm]{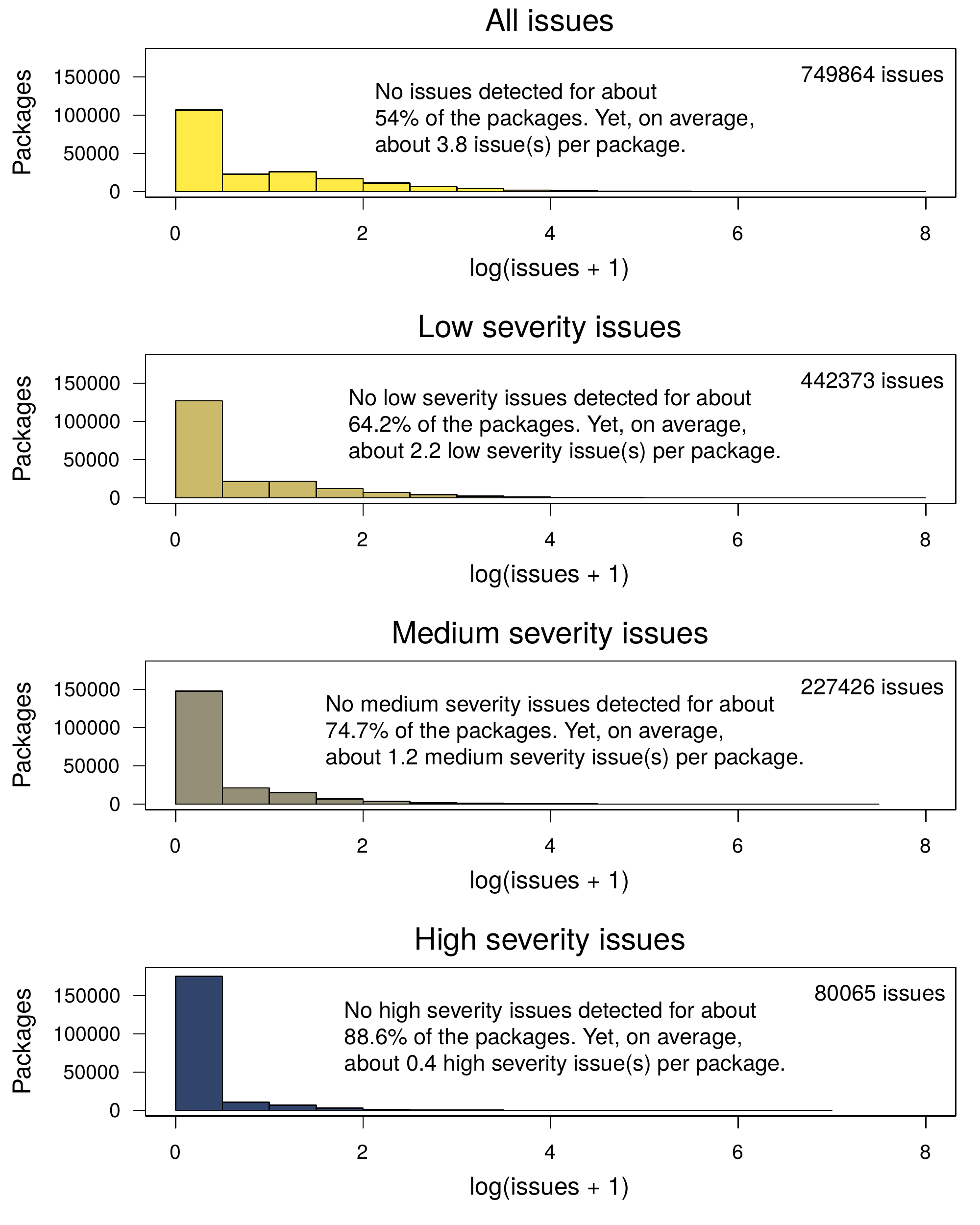}
\caption{Issues Detected}
\label{fig: issues}
\end{figure}

In total, as many as $m = 749,864$ issues were detected for the $n$
packages. Thus, on average, roughly about four security-related issues were
detected across the Python packages. However, these issues seem to concentrate
to particular packages since no issues were detected for a little below the half
of the packages. As can be seen from Fig.~\ref{fig: issues}, the majority of the
issues have a low severity. Medium and high severity issues together account for
about 41\% of the $m$~issues detected. This breakdown according to severity is
not a superficial artifact from the tool used: according to Table~\ref{tab:
  issues}, the tool is not biased toward low-severity issues in
general. Moreover, all four distributions visualized in Fig.~\ref{fig: issues}
have a similar shape. That is, there are many packages with no issues, some
packages with a few issues, and a minority with many issues. Of the 46\% of all
packages with at least one issue, the median number of issues is three and the
75th quartile is~7.

%
%

At the very end of the tail, there are five packages with more than a thousand
detected issues: \texttt{PyGGI}, \texttt{appengine-sdk},
\texttt{genie.libs.ops}, \texttt{pbcore}, and \texttt{genie.libs.parser},
respectively. All have large code bases, which partially explains the large
number of issues detected for these packages. However, interestingly, there
exists variance across the types of issues detected. Of the $2,589$ issues
detected for \texttt{PyGGI}, all are about the ``try-except-pass'' construct
(B110), which, at least without further validation, may be more of a code
smell. However: of the $2,356$ issues detected for \texttt{appengine-sdk}, only
$395$ belong to the generic category; there are also $351$ issues belonging to
the injection category, $500$ issues that are about potential cross-site
scripting, seven issues about potentially insecure use of network protocols, and
so forth and so on. Even the single debug mode issue (B201) is present. Although
these observations are partially explained by the fact that
\texttt{appengine-sdk} embeds a large amount of third-party libraries directly
to its code base, a closer look reveals numerous problematic and potentially
insecure coding practices. These observations motivate
to continue toward examining the issue types more generally.

\begin{figure}[p!]
\centering
\includegraphics[width=\linewidth, height=14cm]{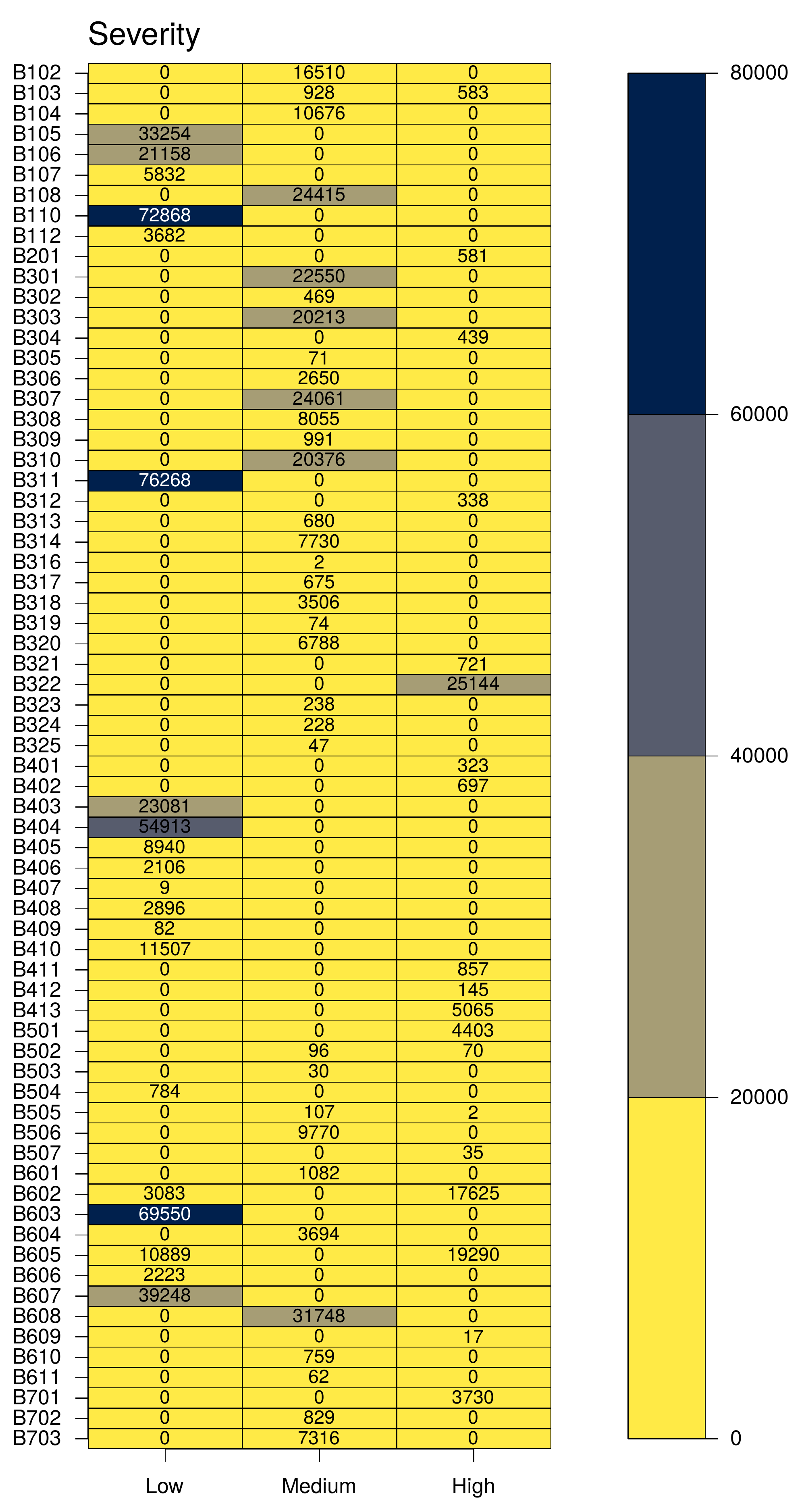}
\caption{Severity of the Issues (frequencies)}
\label{fig: issues severity}
\vspace{20pt}
\includegraphics[width=\linewidth, height=8cm]{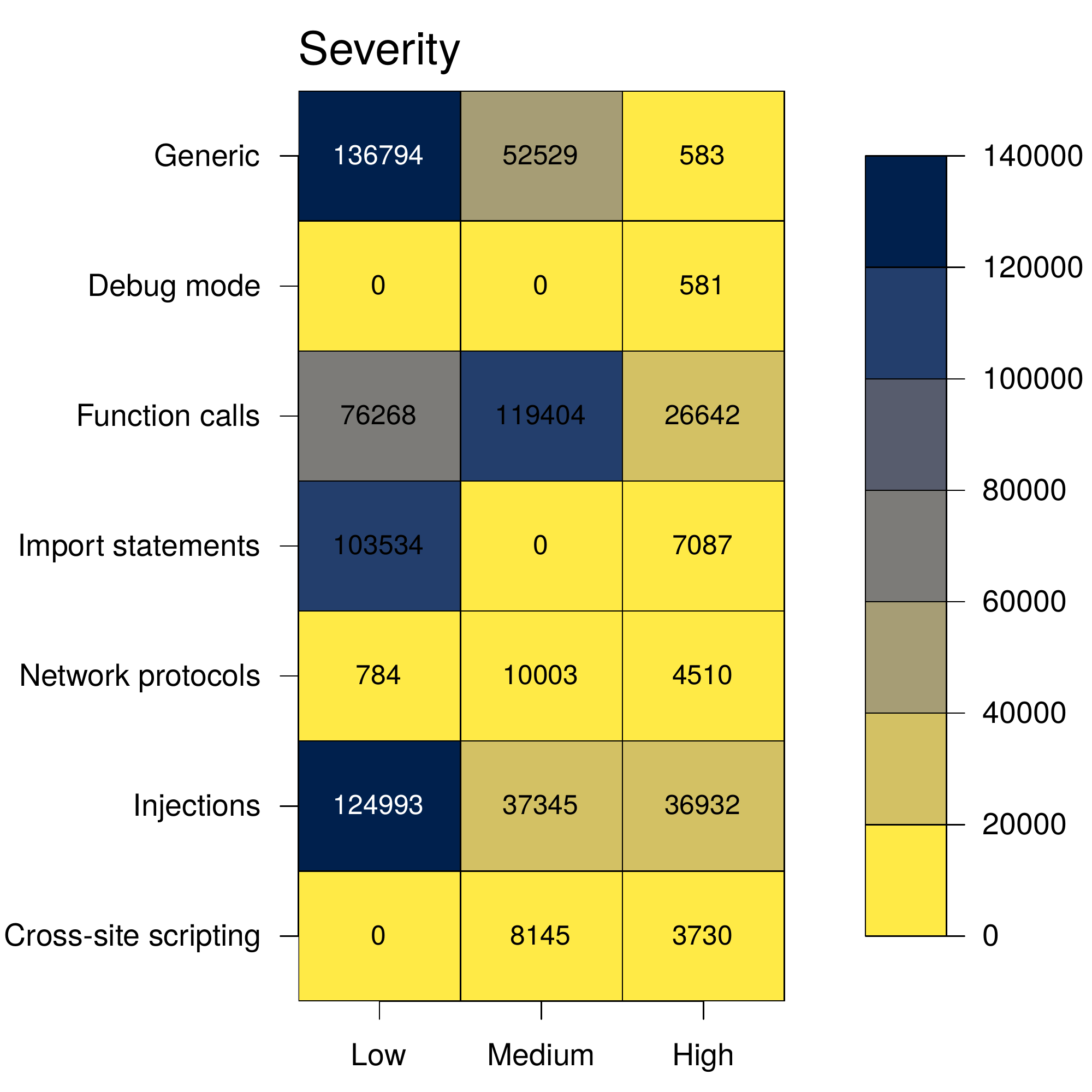}
\caption{Severity of the Issues Across Groups (frequencies)}
\label{fig: groups severity}
\end{figure}

\begin{figure}[p!]
\includegraphics[width=\linewidth, height=14cm]{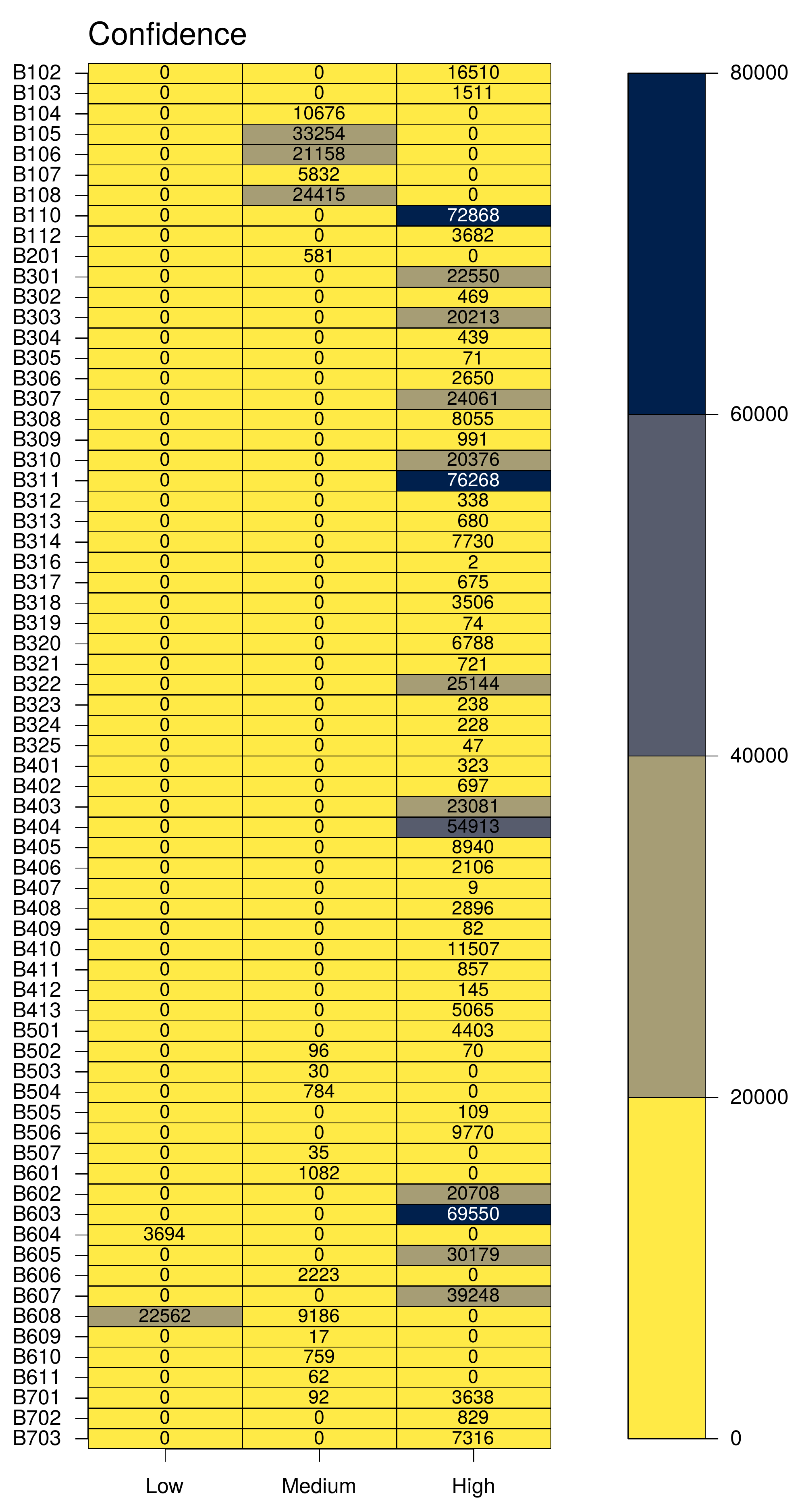}
\caption{Detection Confidence (frequencies)}
\label{fig: issues confidence}
\vspace{20pt}
\includegraphics[width=\linewidth, height=8cm]{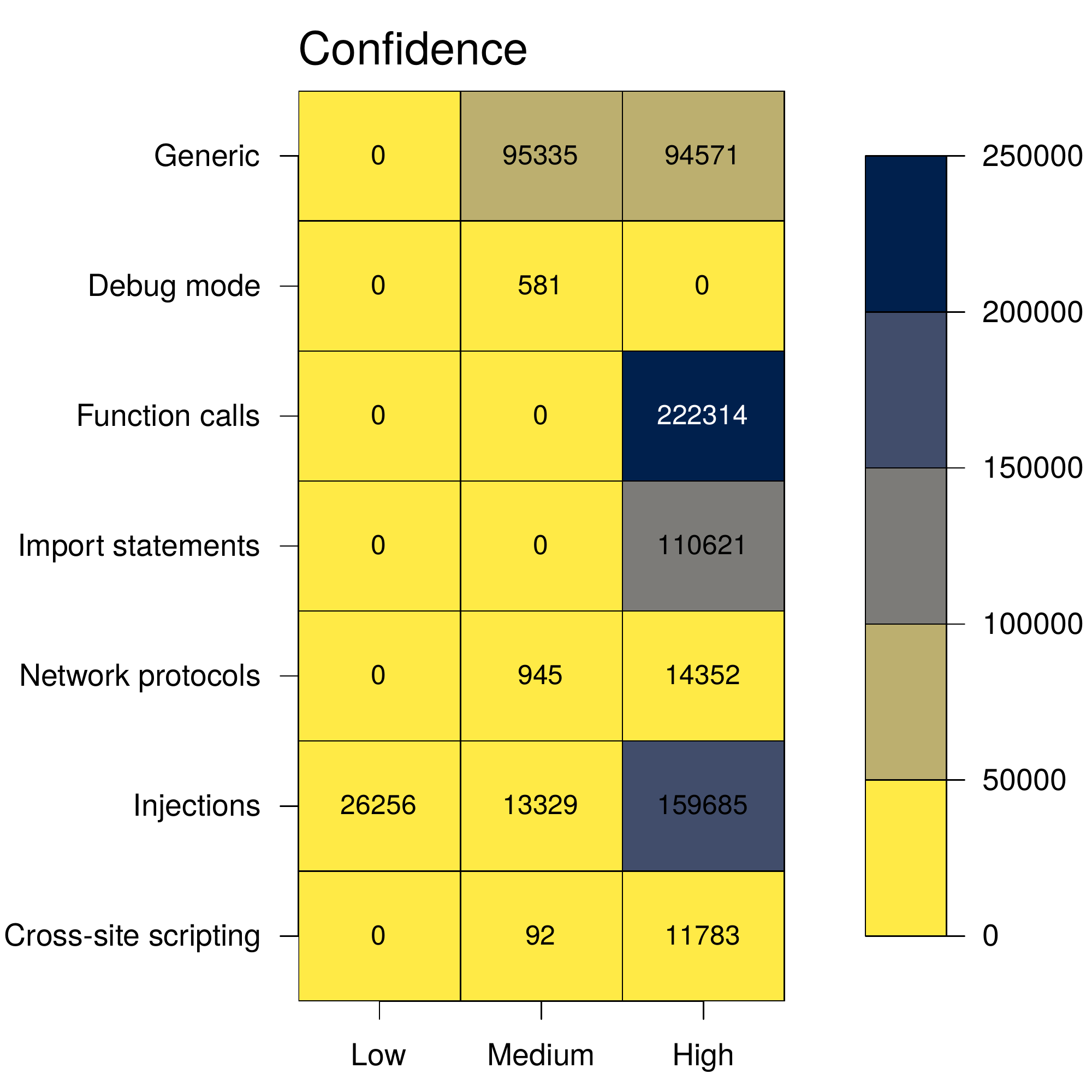}
\caption{Detection Confidence Across Groups (frequencies)}
\label{fig: groups confidence}
\end{figure}

\subsection{Types ($\textmd{RQ}_2$)}

The frequencies of the issues detected across the individual types and the seven categories are shown in Figs.~\ref{fig: issues severity} and \ref{fig: groups severity}. Analogous results are shown in Figs.~\ref{fig: issues confidence} and \ref{fig: groups confidence} for the tool's detection confidence. In general, generic issues (including particularly the already noted B110) and different injections (including particularly those related to the \texttt{subprocess} module) have been the most common ones. Together the generic and injection categories account for over a half (52\%) of all issues detected. However, these are mostly low severity issues. Although the tool's overall self-reported detection confidence is high, the generic and injection categories also gather most of the low and medium detection confidence cases.
%

These results align well with existing observations; command line injections and generally inadequate input validation have been typical to Python applications~\cite{Ruohonen18IWESEP, Rahman19}. In fact, input validation was already one of the kingdoms in the famous ``seven pernicious kingdoms'' of software security issues~\cite{Tsipenyuk05}. Some of the results may relate also to the age of the packages in PyPI. The over twenty thousand issues about insecure hash functions (B303) would be a good example in this regard. Nevertheless, the prevalence of different injections indicate that also Python's standard library is prone to misuse. While all of these issues are noted in the official documentation of the standard library, the issues still frequently appear in PyPI.

\subsection{Software Size ($\textmd{RQ}_3$)}

\begin{figure}[th!b]
\centering
\includegraphics[width=\linewidth, height=4cm]{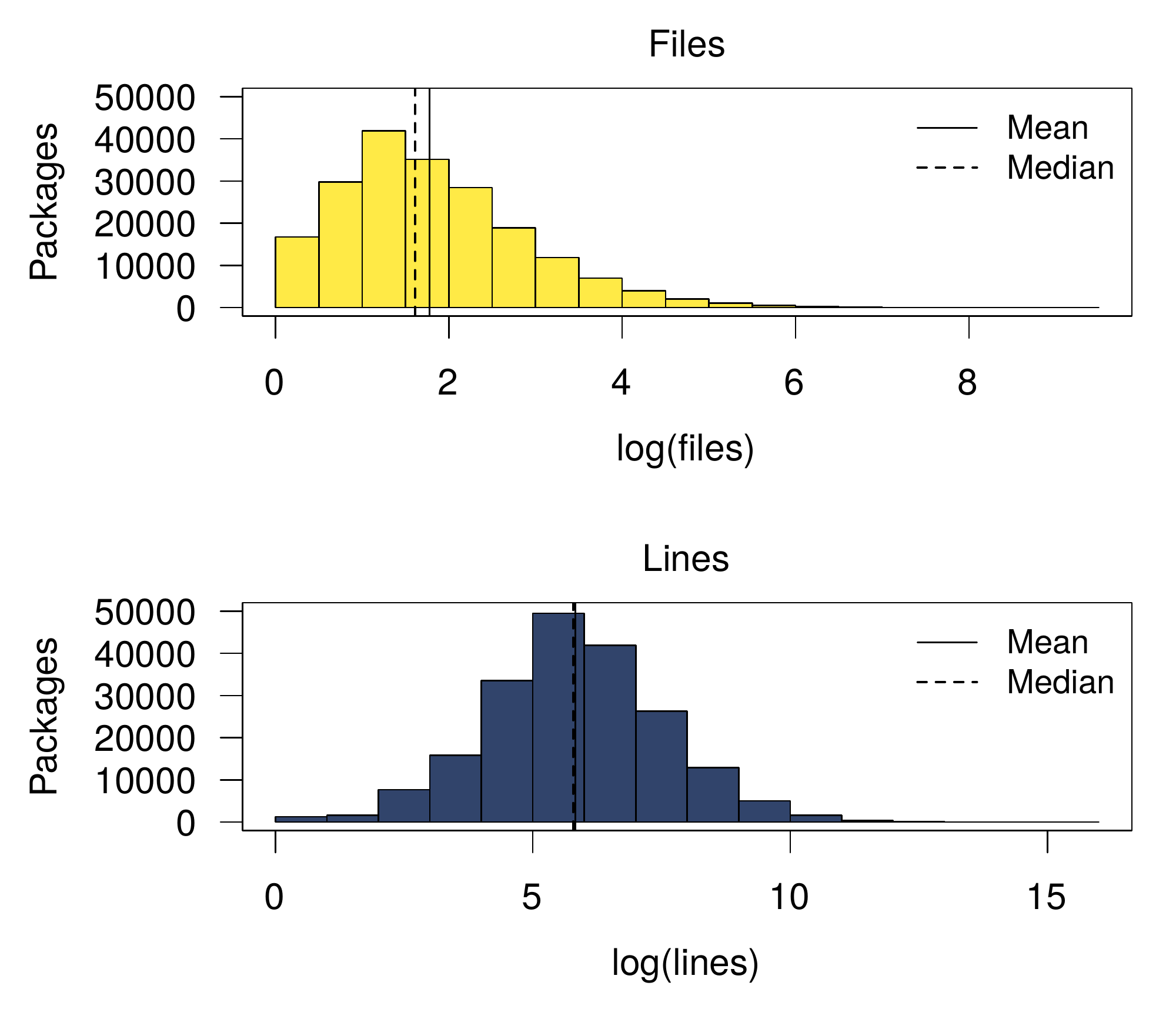}
\caption{Files and Lines of Code}
\label{fig: software size}
\end{figure}

The Python packages in PyPI are generally small. This observation can be seen from Fig.~\ref{fig: software size}, which shows the logarithms of the files and lines scanned. In terms of the files scanned, the largest package is \texttt{alipay-sdk-python} ($9,801$ files); with respect to lines, \texttt{huBarcode} is the largest with its about about $5.9$ million lines of code scanned by Bandit. These two packages are clear outliers, however. For most of the packages, only about three to fifty Python files were scanned. Indirectly, these results indicate that some existing datasets for Python code (such as those used in \cite{Zhifei18} and \cite{Xia18}) are biased upwards toward large projects. In any case, the smallness of the PyPI packages and the shapes of the two distributions in Fig.~\ref{fig: software size} lead to expect that the conditional means are close to the unconditional means (cf.~$\textmd{RQ}_3$). Before examining this prior expectation further, a few points are in order about the two negative binomial regressions (see Subsection~\ref{subsec: methods}) estimated.

The two NB equations \eqref{eq: model 1} and \eqref{eq: model 2} were estimated
with the standard R (MASS) function \texttt{glm.nb}. The estimated dispersion
parameters are nowhere near zero ($\hat{\varphi}_1 \approx 2.999$ and
$\hat{\varphi}_2 \approx 2.381$), which indicates that plain Poisson regressions
would have been inadequate. Although both regression coefficients are positive
and $\hat{\beta_1} > \hat{\beta_2}$, as was expected in Subsection \ref{subsec:
  methods}, the magnitudes of the coefficients are small: $\hat{\beta_1} \approx
0.859$ and $\hat{\beta_2} \approx 0.748$. Already due to the large $n$, these
estimates are expectedly both statistically significant and accurate. The
95\%-level confidence intervals are $[0.852, 0.866]$ for $\hat{\beta_1}$ and
$[0.743, 0.754]$ for the second coefficient estimated.

%

\begin{figure}[th!b]
\centering
\includegraphics[width=\linewidth, height=9cm]{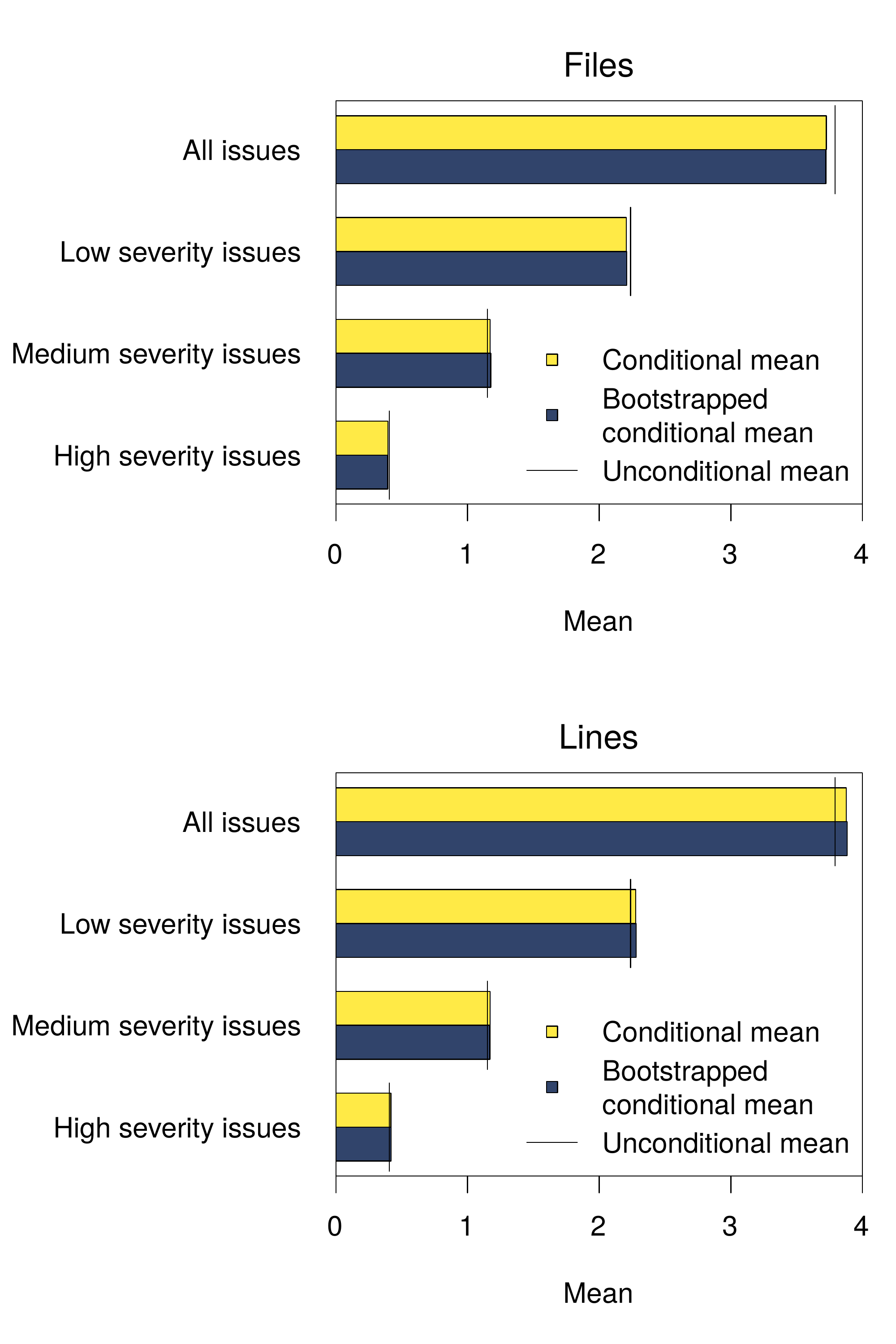}
\caption{Regression Results}
\label{fig: reg nb}
\end{figure}

The small magnitudes of the coefficients indicate relatively modest performance
in general. Indeed, for instance, \text{McFadden}'s~\cite{McFadden74}
pseudo-$R^2$ values are only $0.09$ and $0.14$ for the two models including
files and lines, respectively. These numbers are implicitly reflected in the
final results shown in Fig.~\ref{fig: reg nb}. As can be seen, the unconditional
means are very close to the averaged conditional means computed via \eqref{eq:
  conditional means}. The estimates are also highly accurate; the bootstrapped
estimates are almost equal with the sample estimates in all eight setups.

\section{Discussion}\label{sec: discussion}

\subsection{Threats to Validity}\label{subsec: threats to validity}

Recently, particular attention has been paid to different sampling strategies in empirical software engineering research~\cite{Baltes20}. The dataset used covers practically the whole PyPI, excluding the few packages that had to be excluded due to practical data processing reasons. Of course, there are commercial Python software, other ecosystems for Python packages (such as GitHub), and so forth. However, no attempts were made to infer beyond PyPI.  Generalizability (a.k.a.~\textit{external validity}) is thus not a particular concern for the present work.

But there are concerns regarding the question whether the metrics used truly
measure what is intended to be measured (a.k.a.~\textit{construct
  validity}). All metrics are extremely simple. There are neither interpretation
problems nor theoretical assumptions. Instead, the potential issues are
technical. Even with the detection confidence metric provided by Bandit,
(a)~there are obviously numerous false positives and negatives in the
dataset. One one hand, the paper's large-scale approach makes it difficult to
speculate about the extent of these. On the other hand, if even 0.01\% of the issues were truly vulnerabilities, the amount would still be above seven thousand. To patch this tool-specific limitation in further work, Bandit's detection
accuracy should be evaluated and compared against other static analysis
tools. Given the security-oriented context, the question is not only about
straightforward programming issues. For instance, many static analysis tools are
known to make many mistakes with respect to cryptography~\cite{Braga19}. A~few
further issues are also worth point out about construct validity.

Unlike in some previous studies~\cite{Orru15}, no attempts were made to
distinguish different Python source code files; (b)~the results are based also
on code that many not be executed in production use. Test cases are a good
example. Though, it is not clear whether the additional code constitutes a real
problem; code quality and security issues apply also to test cases. The reverse
also applies: (c)~not all code is covered for some packages. For instance, there
are packages that contain additional archives that were not extracted. Data
files embedded directly to Python code are a good example. As is
well-known~\text{\cite{Aloraini17, Ruohonen18IWESEP}}, packages are often
written with multiple languages; yet, (d)~the results do not cover non-Python
code that may be particularly relevant in terms of security. In general and to
some extent, these construct validity issues are lessened by framing the issues
as code smells~\text{\cite{Rahman19, Zhifei18}}. 

Finally, the assumptions concerning cause and effect (a.k.a.~\textit{internal validity}) should be briefly contemplated. As was noted already in Subsection~\ref{subsec: research questions}, the two simple equations \eqref{eq: model 1} and \eqref{eq: model 2} do not posit meaningful causal relations. Though, the results presented allow to also question whether software size is even a meaningful confounding factor in the context of Python packages. Despite this probable assumption, (e) only a snapshot was used; the results are limited to one point in time. The omitted dynamics are a potential problem for the regression analysis, but the lack of attention paid to software evolution can be seen also as a problem of external validity.

\subsection{Further Work}

Some fruitful paths for further work can be mentioned. Patching some of the limitations discussed is the obvious starting points: further large-scale studies are needed to compare different static analysis tools for Python; further software metrics could be collected to make the simple predictive approach more interesting theoretically; longitudinal analysis is required for assessing the robustness of the effects reported; and so forth. The last point about longitudinal research setups is also important for knowing whether things are improving or getting worse. Although many individual projects continuously fix issues in their source code, it may be that such fixing does not improve the situation in the whole ecosystem. To this end, it would be interesting to examine whether automated solutions based on static analysis could be incorporated into the PyPI infrastructure. After all, automation and integration to continuous integration or other related software systems have often been seen important for making practical advances with static analysis~\cite{Beller16, ChessMcGraw04}. A simple option would be to scan a package via a static analysis tool each time the package is uploaded to PyPI, and then, upon request, display the tool's output for a user, possibly via the \texttt{pip} package manager.

As for further empirical research, it would be interesting to examine whether
the results differ between different types of Python packages. To this end,
there exists previous work regarding the so-called ``GitHub stars''
\cite{Borges18}.  In PyPI there are also ``classifiers''
\cite{PyPI20b} that could be used for the comparison. 


\section{Conclusion}\label{sec: conclusion}

This paper examined a snapshot of almost all packages archived to PyPI. Three research questions were asked. To briefly summarize the answers to these, security issues are common in PyPI packages. At least one issue was detected for a little below half of the packages in the dataset; though, the majority of the issues observed are of low severity ($\textmd{RQ}_1$). The issue types show no big surprises: different code injections and generally inadequate input validation characterize most of the issues~($\textmd{RQ}_2$). Finally, the generally small size of the Python packages in PyPI implies that basic code size metrics do not predict well the per-package amount of issues detected~($\textmd{RQ}_3$).

\balance
\bibliographystyle{abbrv}

\end{document}